\documentclass[twocolumn,prl,superscriptaddress,showpacs]{revtex4}

\usepackage{amsmath,amsfonts,amssymb,bm}
\usepackage{dcolumn}
\usepackage[final]{graphicx}
\usepackage{bm}

\newcommand{\proj}[2]{|#1\rangle\langle#2|}

\begin{document}

\title{Quantum control of electron--phonon scatterings in artificial atoms}

\author{Ulrich Hohenester}\email{ulrich.hohenester@uni-graz.at}
\affiliation{Institut f\"ur Theoretische Physik,
  Karl--Franzens--Universit\"at Graz, Universit\"atsplatz 5,
  8010 Graz, Austria}

\author{Georg Stadler}
\affiliation{Institut f\"ur Mathematik, Karl--Franzens--Universit\"at Graz, 
Heinrichstra\ss e 36, 8010 Graz, Austria}

\date{January 19, 2004}

\begin{abstract}

The phonon-induced dephasing dynamics in optically excited semiconductor quantum dots is studied within the frameworks of the independent Boson model and optimal control. We show that appropriate tailoring of laser pulses allows a complete control of the optical excitation despite the phonon dephasing, a finding in marked contrast to other environment couplings. 

\end{abstract}



\maketitle

Ultrafast spectroscopy allows to investigate the fundamental interaction mechanisms in semiconductors by means of optical excitation of electrons and holes~\cite{shah:96,rossi:02}, which consecutively relax through mutual scatterings~\cite{huber:01} or interactions with the lattice~\cite{hase:03}. While in bulk semiconductors the scattering times are completely determined by the genuine solid-state interactions, appropriate tailoring of the carrier density of states in lower-dimensional semiconductor structures makes possible a modification of the scattering characteristics. An extreme case is found in semiconductor quantum dots~\cite{bimberg:98,hawrylak:98}, often referred to as {\em artificial atoms},\/ where the optical excitation of an electron-hole pair in the state of lowest energy causes the deformation of the surrounding lattice but relaxation is completely inhibited because of the atomic-like carrier density of states. In coherent optical spectroscopy \cite{shah:96,rossi:02}, which is sensitive to the optically induced coherence, this partial transfer of quantum coherence from the electron-hole state to the lattice degrees of freedom, i.e., phonons, results in {\em dephasing}\/~\cite{borri:01}. The coupled dot-phonon system is conveniently described within the independent Boson model~\cite{mahan:81}, which is exactly solvable for laser pulses $\delta$-like in time~\cite{krummheuer:02,vagov:02,jacak:03} whereas approximate description schemes have to be employed for laser pulses of finite duration. This has recently been accomplished within a density-matrix approach by F\"orstner et al.~\cite{foerstner:03a}, who reported a surprisingly large impact of phonon-assisted dephasing on the coherent optical response. Apparently, this constitutes a serious drawback for all-optical quantum information applications in quantum dots, which have recently received considerable interest~\cite{barenco:95,troiani.prb:00,biolatti:00,zrenner:02,li:03}.

It should, however, be noted that, contrary to other decoherence channels in solids where the system's wavefunction acquires an uncontrollable phase through environment coupling~\cite{zurek:03}, in the independent Boson model the loss of phase coherence is due to the coupling of the electron-hole state to an ensemble of harmonic oscillators which all evolve with a coherent time evolution but different phase. This results in destructive interference and dephasing, and thus spoils the direct applicability of coherent carrier control. On the other hand, the coherent nature of the state-vector evolution suggests that more refined control strategies might allow to suppress dephasing losses. To address this problem, in this letter we examine phonon-assisted dephasing within the framework of {\em optimal control}\/ \cite{rabitz:00,peirce:88,borzi.pra:02} aiming at a most efficient control strategy to channel the system's wavefunction through a sequence of given states. We find that appropriate tailoring of laser pulses allows to promote the system from the ground state through a sequence of excited states back to the ground state {\em without suffering significant dephasing losses}.\/ Despite the widespread use of the independent Boson model, e.g., for the description of optical properties of localized states in solids or M\"o\ss bauer spectroscopy, to our best knowledge no such control strategy for suppression of environment losses has hitherto been reported in the literature. Our results thus not only reinforce quantum dots as viable candidates for quantum-information processing devices, but also suggest related schemes for the control of impurity or defect states in solids, e.g., in the context of ultraslow light propagation \cite{turukhin:02} or creation of multiparticle entanglement \cite{bao:03}.

In our theoretical approach we follow Refs.~\cite{krummheuer:02,vagov:02,foerstner:03a} and start with the usual independent Boson Hamiltonian. We describe the dot states in terms of a generic two-level system, with ground state $0$ and excited state $x$, assuming a negligible contribution of excited exciton states due to the typically large energy splittings of several tens of meV \cite{bimberg:98} and of biexcitons, which, in optical experiments, can be achieved through appropriate polarization filtering. This two-level system is coupled to a reservoir of harmonic oscillators such that the interaction only occurs when the system is in the upper state \cite{mahan:81}:

\begin{figure}
\caption{Results of our calculations with Gaussian $2\pi$ (dashed lines) and optimal-control (solid lines) laser pulses and for: (a,d) zero temperature; (b,e) zero temperature and an electron-phonon coupling enhanced by a factor of $\sqrt 3$; and (c,e) finite temperature $T=\omega_c\sim 10$ K. The upper graphs (a--c) show $|\Omega(t)|$, the lower ones (d--f) the time evolution of $u_3(t)$, and the insets the trajectories of the Bloch vector $\bm u(t)$. For the Gaussian $2\pi$-pulse Rabi flopping occurs but is damped due to electron-phonon interactions. For the optimal control dephasing losses are completely suppressed, and the the system passes through the desired states of $\hat{\bm e}_3$ at time zero and $-\hat{\bm e}_3$ at $T$. 
}
\end{figure}

\begin{figure}
\caption{Time evolution of the different phonon occupancies in the excited state $\sum_\lambda |\, \langle x|a_\lambda|\psi(t)\rangle\,|^2\,\delta(\omega-\omega_\lambda)$ as a function of phonon energy $\omega$ and for the (a) Gaussian and (b) optimal control shown in Fig.~1(b). 
}
\end{figure}

\begin{eqnarray}\label{eq:spin-boson}
H&=&\sum_\lambda g_\lambda\, (a_\lambda+a_\lambda^\dagger)\,\proj x x
+\sum_\lambda \omega_\lambda\, a_\lambda^\dagger a_\lambda\nonumber\\
&-&\mbox{$\frac 1 2$}\bigl(\,\Omega\,\proj x 0 + \Omega^*\,\proj 0 x \,\bigr)\,.
\end{eqnarray}

\noindent Here, the bosonic degrees of freedom $\lambda$ with energy $\omega_\lambda$ are described by the bosonic field operators $a_\lambda$ and $a_\lambda^\dagger$, and $g_\lambda$ is the coupling constant between $x$ and $\lambda$. In Eq.~(\ref{eq:spin-boson}) we have neglected in accordance to Refs.~\cite{krummheuer:02,vagov:02,jacak:03,foerstner:03a} higher-order phonon processes, e.g., anharmonic decay or point-defect scatterings, which is a well controlled approximation in view of the long calculated \cite{tamura:85} and measured \cite{bartels:98,kent:02} acoustic-phonon lifetimes. We finally describe the light-matter coupling within the usual dipole and rotating-wave approximations~\cite{mandel:95} with $\Omega=\bm\mu\,\bm{\mathcal{E}}$ the Rabi frequency, $\bm\mu$ the dipole moment of the two-level system, and $\bm{\mathcal{E}}$ the envelope of the external laser control. Similar to Ref.~\cite{foerstner:03a} we consider a spherical dot model and acoustic deformation potential interactions $g_q=(q/2\rho c)^\frac 1 2(D_e-D_h)\exp(-q^2\sigma^2/4)$ as the only coupling mechanism, with $q$ the phonon wavevector, $\rho$ the mass density, $c$ the longitudinal sound velocity, $D_e$ and $D_h$ the deformation potentials for electrons and holes, respectively, and $\sigma$ the carrier localization length. We assume $\sigma=5$ nm and use material parameters representative for GaAs \cite{foerstner:03a}. In the following length is measured in units of $\sigma$, energy in units of $\omega_c=c/\sigma\sim 0.7$ meV, and time in units of $\omega_c^{-1}\sim 1$ ps.

{\em Zero temperature}.---At zero temperature the solution of the independent boson model \eqref{eq:spin-boson} reduces to that of the Schr\"odinger equation $i|\dot\psi\rangle=H|\psi\rangle$, which we numerically solve by expanding $|\psi\rangle$ in the basis of $|0\rangle$, $|x\rangle$, $a_\lambda^\dagger|0\rangle$, and $a_\lambda^\dagger|x\rangle$. The control $\Omega(t)$ is assumed to have a Gaussian envelope with a full-width of half maximum of $5\,\omega_c^{-1}$ and area $2\pi$, and to be tuned to the polaron ground state frequency $\Delta=-\sum_\lambda (g_\lambda)^2/\omega_\lambda$ \cite{mahan:81}. To describe the two-level system we introduce the {\em Bloch vector}\/ $\bm u=\langle\bm\sigma\rangle$, with $\sigma_i$ the usual Pauli matrices \cite{mandel:95}, where $u_1$ and $u_2$ account for the real and imaginary part of the interband polarization, respectively, and $u_3$ gives the population difference between $0$ and $x$. While in absence of phonon coupling $\Omega(t)$ would simply rotate $\bm u$ from $-\hat{\bm e}_3$ through a series of excited states back to $-\hat{\bm e}_3$, the phonon coupling of Eq.~\eqref{eq:spin-boson} entangles the two-level system with the lattice degrees of freedom and leads to dephasing. This can be seen in Fig.~1(d) which shows that after the action of the Gaussian $2\pi$-pulse the Bloch vector remains in an excited state. From the inset of Fig.~1(d), and even more clearly from Fig.~1(e) showing results for a phonon coupling enhanced by a factor of $\sqrt 3$, we observe that the final deviation of $\bm u$ from the initial state $-\hat{\bm e}_3$ is not due to an incomplete rotation of $\bm u$ but to a loss of norm of $\|\bm u\|$, i.e., {\em the system has suffered dephasing losses}.\/

In the following we shall address the question whether such losses are inherent to the system under investigation or can be suppressed by more sophisticated control strategies. To this end, we quantify the objective of the control through the cost function $J(\bm u,\Omega)=\frac 1 2(\int_{-T}^T dt\,\beta(t)|\bm u(t)-\hat{\bm e}_3|^2+|\bm u(T)+\hat{\bm e}_3|^2+\alpha\int_{-T}^T dt\,|\Omega(t)|^2)$, with $\beta$ a Gaussian centered at time zero with a narrow half-width of full maximum of $0.1\,\omega_c^{-1}$ and $\alpha=10^{-5}$ a small constant, i.e, we are seeking for solutions where $\bm u$ passes through $\hat{\bm e}_3$ at time zero and goes back to $-\hat{\bm e}_3$ at $T$, and the last term in $J(\bm u,\Omega)$ accounts for the limited laser resources and is needed to make the optimal-control problem well posed. 
To determine the optimal control we have to minimize $J(\bm u,\Omega)$ subject to the constraint that $\psi$ fulfills Schr\"odinger's equation with the initial condition $|\psi(-T)\rangle=|0\rangle$, which is done in accordance to Refs.~\cite{peirce:88,borzi.pra:02} by introducing Lagrange multipliers for the constraints and 
utilizing that the Lagrange functional admits a stationary point at the solution. The resulting optimality system is solved by use of a gradient-related method including linesearch \cite{borzi.pra:02}, which requires to integrate the state equation for $\psi$ forwards in time and an adjoint equation for the dual variable $\tilde\psi$ backwards in time. The solutions $\psi$ and $\tilde\psi$ provide us with an improved $\Omega(t)$, and allow for an iterative minimization of $J(\bm u,\psi)$ that converges after typically a few hundred iterations to the {\em optimal control}.\/ Results of our optimal-control calculations are shown in Fig.~1. Most remarkably, we can indeed obtain a control field for which $\bm u(t)$ passes through the desired states of $\hat{\bm e}_3$ at time zero and $-\hat{\bm e}_3$ at $T$. {\em Thus, appropriate pulse shaping allows to fully control the two-level system even in presence of phonon couplings}.\/

We emphasize that, contrary to the quantum ``bang-bang'' control reported by Viola and Lloyd \cite{viola:98} where the system is constantly flipped to suppress decoherence, in Fig.~1 the Bloch vector is rotated slowly such that all oscillators can follow the system almost adiabatically irrespective of $\omega_\lambda$. This is shown clearly in Fig.~2 where we plot the time evolution of the different phonon occupancies in the excited state $\sum_\lambda |\, \langle x|a_\lambda|\psi(t)\rangle\,|^2\,\delta(\omega-\omega_\lambda)$. A comparison of the Gaussian and optimal control strategies of Figs.~2(a) and (b) reveals that only in the latter case all low-energy phonon modes can follow $\bm u(t)$ and become de-excited at the end of the pulse. In a sense, this finding is reminiscent of self-induced transparency \cite{mccall:67,mandel:95,panzarini.prb:02} where an intense laser pulse propagating in a medium of inhomogeneously broadened two-level systems acquires a pulse shape which drives all systems from the groundstate through a sequence of excited states back to the groundstate irrespective of their detuning.

{\em Finite temperature}.---We next address the question whether our findings prevail in case of finite temperature. Similar to Ref.~\cite{foerstner:03a}, we use a density-matrix approach with $\bm u=\langle\bm\sigma\rangle$ the Bloch vector, $s_\lambda=\langle a_\lambda\rangle$ the coherent phonon amplitude, and $\bm u_\lambda=\langle \bm\sigma(a_\lambda-s_\lambda)\rangle$ the phonon-assisted density matrix as dynamic variables. Their time evolution is governed by \cite{unpublished}:

\begin{subequations}\label{eq:density-matrix}
\begin{eqnarray}
\dot{\bm u}&=&\bm\Omega\times \bm u+2\sum_\lambda g_\lambda\,
\hat{\bm e}_3\times\Re e(\bm u_\lambda)\\
\dot s_\lambda &=& -i\,\omega_\lambda\,s_\lambda - \mbox{$\frac i 2$}\, g_\lambda\, (1+u_3)\\
\dot{\bm u}_\lambda&=&\bm\Omega\times \bm u_\lambda-i\,\omega_\lambda\,\bm u_\lambda \nonumber\\
&+&g_\lambda\, (n_\lambda+\mbox{$\frac 1 2$})\, \hat{\bm e}_3\times \bm u
+\mbox{$\frac i 2$}\, g_\lambda(u_3\,\bm u-\hat{\bm e}_3)\,,
\end{eqnarray}
\end{subequations}

\noindent with the initial conditions $\bm u(-T)=-\hat{\bm e}_3$, $s_\lambda(-T)=0$, and $\bm u_\lambda(-T)=0$. Here, $n_\lambda=\langle (a_\lambda^\dagger-s_\lambda^*)(a_\lambda-s_\lambda)\rangle$ is the occupation of incoherent phonons, which we shall approximate by a Bose-Einstein distribution~\cite{foerstner:03a}, and we have introduced the abbreviation $\bm\Omega=-\Re e(\Omega)\,\hat{\bm e}_1+\Im m(\Omega)\,\hat{\bm e}_2+2\,\Re e\sum_\lambda g_\lambda\, s_\lambda\,\hat{\bm e}_3$. We again use the method of Lagrange multipliers to minimize $J(\bm u,\Omega)$ subject to Eq.~\eqref{eq:density-matrix}, and obtain the adjoint equations \cite{unpublished}:

\begin{widetext}

\begin{subequations}\label{eq:adjoint}
\begin{eqnarray}
\dot{\tilde{\bm u}}&=& \bm\Omega\times \tilde{\bm u} +\sum_\lambda g_\lambda
\biggl((n_\lambda+\mbox{$\frac 1 2$})\,\hat{\bm e_3}\times\Re e(\tilde{\bm u}_\lambda)+\mbox{$\frac 1 2$}\, \Im m\Bigl(
(\tilde s_\lambda-\bm u\,\tilde{\bm u}_\lambda)\,\hat{\bm e}_3
-u_3\,\tilde{\bm u}_\lambda\Bigr)
\biggr)+\beta(t)(\bm u-\hat{\bm e}_3)\nonumber \\  &&\\
\dot{\tilde s}_\lambda &=& -i\omega_\lambda\tilde s_\lambda+
2\,g_\lambda\,\biggl((\tilde{\bm u}\times \bm u)\,\hat{\bm e}_3+\Re e\sum_{\lambda'}
(\tilde{\bm u}_{\lambda'}^*\times {\bm u}_{\lambda'})\,\hat{\bm e}_3\biggr)\\
\dot{\tilde{\bm u}}_\lambda &=& \bm\Omega\times\tilde{\bm u}_\lambda-
i\omega_\lambda \tilde{\bm u}_\lambda+2\,g_\lambda\, \hat{\bm e}_3\times 
\tilde{\bm u}\,,
\end{eqnarray}
\end{subequations}

\end{widetext}

\noindent with terminal conditions $\tilde{\bm u}(T)=-\hat{\bm e}_3-\bm u(T)$, $\tilde s_\lambda(T)=0$, and $\tilde{\bm u}_\lambda(T)=0$. Eqs.~\eqref{eq:density-matrix} and \eqref{eq:adjoint} together with

\begin{equation}\label{eq:control}
\Omega=\frac 1\alpha\biggl((\tilde{\bm u}\times \bm u)+
\Re e\sum_\lambda (\tilde{\bm u}_\lambda^*\times \bm u_\lambda)
\biggr)\,(\hat{\bm e}_1-i\hat{\bm e}_2)
\end{equation}

\noindent form the optimality system. Similar to Ref.~\cite{borzi.pra:02} it is solved iteratively through integration of Eq.~\eqref{eq:density-matrix} forwards and Eq.~\eqref{eq:adjoint} backwards in time, and computing an improved control by use of Eq.~\eqref{eq:control}.

At zero temperature, i.e., for $n_\lambda=0$, the solutions of Eq.~\eqref{eq:density-matrix} almost coincide with the corresponding wavefunction results shown in Figs.~1(d,e). With increasing temperature it is still possible to find optimal-control fields where dephasing losses are strongly suppressed, but the control strategy gradually changes to that reported in Fig.~1(f) for a temperature of $T=\omega_c\sim 10$ K. Here, the transition time of $\bm u$ from $-\hat{\bm e}_3$ to $\hat{\bm e}_3$ is minimized and consequently dephasing losses, which only occur for finite values of $u_1$ and $u_2$, are suppressed. However, as compared to the low-temperature case this control strategy is much more trivial, and could have been guessed from more simple considerations.

In conclusion, we have studied the phonon-induced dephasing dynamics in optically excited semiconductor quantum dots within the frameworks of the independent Boson model and optimal control. We have shown that appropriate tailoring of laser pulses allows to control the dot states without suffering significant dephasing losses, not only at the lowest but, though exceedingly difficult, also at elevated temperatures. The requirements for such laser-pulse shaping are well within the possibilities of presentday technology \cite{rabitz:00}. To highlight the applicability of quantum control, in this work we have focused on laser pulses with durations of a few picoseconds where the effects of dephasing losses are most pronounced. For other control objectives it might be advantageous to use shorter or longer laser pulses, for which control becomes substantially simplified, or to rely on more advanced control strategies such as, e.g., stimulated Raman adiabatic passage \cite{bergmann:98,borzi.pra:02}. Besides their importance for future quantum-information processing applications, our findings might be also useful to address more fundamental questions regarding the nature of scatterings in solids. Contrary to higher-dimensional semiconductors, where scatterings occur on a sub-picosecond timescale \cite{shah:96,rossi:02,huber:01,hase:03}, within the present scheme the atomic-like density of states leads to a significant slow-down of electron-phonon scatterings and allows their manipulation through an external control. This situation resembles that of controlled collisions of optically trapped atoms which are used to establish multi-particle entanglement \cite{mandel:03}. In a similar fashion, one might envision that controlled electron-phonon scatterings in semiconductors, i.e., genuine solid-state interaction channels, might open a way for tunable interdot interactions.

We are grateful to Alfio~Borz\`{\i} for most helpful discussions. This work has been supported in part by the {\em Fonds zur F\"orderung der wissenschaftlichen Forschung}\/ (FWF) under SRC 03 ''Optimization and Control'' and project P15752--N08.


\begin{thebibliography}{33}
\expandafter\ifx\csname natexlab\endcsname\relax\def\natexlab#1{#1}\fi
\expandafter\ifx\csname bibnamefont\endcsname\relax
  \def\bibnamefont#1{#1}\fi
\expandafter\ifx\csname bibfnamefont\endcsname\relax
  \def\bibfnamefont#1{#1}\fi
\expandafter\ifx\csname citenamefont\endcsname\relax
  \def\citenamefont#1{#1}\fi
\expandafter\ifx\csname url\endcsname\relax
  \def\url#1{\texttt{#1}}\fi
\expandafter\ifx\csname urlprefix\endcsname\relax\def\urlprefix{URL }\fi
\providecommand{\bibinfo}[2]{#2}
\providecommand{\eprint}[2][]{\url{#2}}

\bibitem[{\citenamefont{Shah}(1996)}]{shah:96}
\bibinfo{author}{\bibfnamefont{J.}~\bibnamefont{Shah}},
  \emph{\bibinfo{title}{Ultrafast Spectroscopy of Semiconductors and
  Semiconductor Nanostructures}} (\bibinfo{publisher}{Springer},
  \bibinfo{address}{Berlin}, \bibinfo{year}{1996}).

\bibitem[{\citenamefont{Rossi and Kuhn}(2002)}]{rossi:02}
\bibinfo{author}{\bibfnamefont{F.}~\bibnamefont{Rossi}} \bibnamefont{and}
  \bibinfo{author}{\bibfnamefont{T.}~\bibnamefont{Kuhn}},
  \bibinfo{journal}{Rev. Mod. Phys.} \textbf{\bibinfo{volume}{74}},
  \bibinfo{pages}{895} (\bibinfo{year}{2002}).

\bibitem[{\citenamefont{Huber et~al.}(2001)\citenamefont{Huber, Tauser,
  Brodschelm, Bichler, Abstreiter, and Leitenstorfer}}]{huber:01}
\bibinfo{author}{\bibfnamefont{R.}~\bibnamefont{Huber}},
  \bibinfo{author}{\bibfnamefont{F.}~\bibnamefont{Tauser}},
  \bibinfo{author}{\bibfnamefont{A.}~\bibnamefont{Brodschelm}},
  \bibinfo{author}{\bibfnamefont{M.}~\bibnamefont{Bichler}},
  \bibinfo{author}{\bibfnamefont{G.}~\bibnamefont{Abstreiter}},
  \bibnamefont{and}
  \bibinfo{author}{\bibfnamefont{A.}~\bibnamefont{Leitenstorfer}},
  \bibinfo{journal}{Nature} \textbf{\bibinfo{volume}{414}},
  \bibinfo{pages}{286} (\bibinfo{year}{2001}).

\bibitem[{\citenamefont{Hase et~al.}(2003)\citenamefont{Hase, Kitajima,
  Constantinescu, and Petek}}]{hase:03}
\bibinfo{author}{\bibfnamefont{M.}~\bibnamefont{Hase}},
  \bibinfo{author}{\bibfnamefont{M.}~\bibnamefont{Kitajima}},
  \bibinfo{author}{\bibfnamefont{A.~M.} \bibnamefont{Constantinescu}},
  \bibnamefont{and} \bibinfo{author}{\bibfnamefont{H.}~\bibnamefont{Petek}},
  \bibinfo{journal}{Nature} \textbf{\bibinfo{volume}{426}}, \bibinfo{pages}{51}
  (\bibinfo{year}{2003}).

\bibitem[{\citenamefont{Bimberg et~al.}(1998)\citenamefont{Bimberg, Grundmann,
  and Ledentsov}}]{bimberg:98}
\bibinfo{author}{\bibfnamefont{D.}~\bibnamefont{Bimberg}},
  \bibinfo{author}{\bibfnamefont{M.}~\bibnamefont{Grundmann}},
  \bibnamefont{and}
  \bibinfo{author}{\bibfnamefont{N.}~\bibnamefont{Ledentsov}},
  \emph{\bibinfo{title}{Quantum dot heterostructures}}
  (\bibinfo{publisher}{{J}ohn {W}iley}, \bibinfo{address}{New York},
  \bibinfo{year}{1998}).

\bibitem[{\citenamefont{Jacak et~al.}(1998)\citenamefont{Jacak, Hawrylak, and
  Wojs}}]{hawrylak:98}
\bibinfo{author}{\bibfnamefont{L.}~\bibnamefont{Jacak}},
  \bibinfo{author}{\bibfnamefont{P.}~\bibnamefont{Hawrylak}}, \bibnamefont{and}
  \bibinfo{author}{\bibfnamefont{A.}~\bibnamefont{Wojs}},
  \emph{\bibinfo{title}{Quantum Dots}} (\bibinfo{publisher}{Springer},
  \bibinfo{address}{Berlin}, \bibinfo{year}{1998}).

\bibitem[{\citenamefont{Borri et~al.}(2001)\citenamefont{Borri, Langbein,
  Schneider, Woggon, Sellin, Ouyang, and Bimberg}}]{borri:01}
\bibinfo{author}{\bibfnamefont{P.}~\bibnamefont{Borri}},
  \bibinfo{author}{\bibfnamefont{W.}~\bibnamefont{Langbein}},
  \bibinfo{author}{\bibfnamefont{S.}~\bibnamefont{Schneider}},
  \bibinfo{author}{\bibfnamefont{U.}~\bibnamefont{Woggon}},
  \bibinfo{author}{\bibfnamefont{R.~L.} \bibnamefont{Sellin}},
  \bibinfo{author}{\bibfnamefont{D.}~\bibnamefont{Ouyang}}, \bibnamefont{and}
  \bibinfo{author}{\bibfnamefont{D.}~\bibnamefont{Bimberg}},
  \bibinfo{journal}{Phys. Rev. Lett.} \textbf{\bibinfo{volume}{87}},
  \bibinfo{pages}{157401} (\bibinfo{year}{2001}).

\bibitem[{\citenamefont{Mahan}(1981)}]{mahan:81}
\bibinfo{author}{\bibfnamefont{G.~D.} \bibnamefont{Mahan}},
  \emph{\bibinfo{title}{Many-particle physics}} (\bibinfo{publisher}{Plenum},
  \bibinfo{address}{New York}, \bibinfo{year}{1981}).

\bibitem[{\citenamefont{Krummheuer et~al.}(2002)\citenamefont{Krummheuer, Axt,
  and Kuhn}}]{krummheuer:02}
\bibinfo{author}{\bibfnamefont{B.}~\bibnamefont{Krummheuer}},
  \bibinfo{author}{\bibfnamefont{V.~M.} \bibnamefont{Axt}}, \bibnamefont{and}
  \bibinfo{author}{\bibfnamefont{T.}~\bibnamefont{Kuhn}},
  \bibinfo{journal}{Phys. Rev. B} \textbf{\bibinfo{volume}{65}},
  \bibinfo{pages}{195313} (\bibinfo{year}{2002}).

\bibitem[{\citenamefont{Vagov et~al.}(2002)\citenamefont{Vagov, Axt, and
  Kuhn}}]{vagov:02}
\bibinfo{author}{\bibfnamefont{A.}~\bibnamefont{Vagov}},
  \bibinfo{author}{\bibfnamefont{V.~M.} \bibnamefont{Axt}}, \bibnamefont{and}
  \bibinfo{author}{\bibfnamefont{T.}~\bibnamefont{Kuhn}},
  \bibinfo{journal}{Phys. Rev. B} \textbf{\bibinfo{volume}{66}},
  \bibinfo{pages}{165312} (\bibinfo{year}{2002}).

\bibitem[{\citenamefont{Jacak et~al.}(2003)\citenamefont{Jacak, Machnikowski,
  Kransnyj, and Zoller}}]{jacak:03}
\bibinfo{author}{\bibfnamefont{L.}~\bibnamefont{Jacak}},
  \bibinfo{author}{\bibfnamefont{P.}~\bibnamefont{Machnikowski}},
  \bibinfo{author}{\bibfnamefont{J.}~\bibnamefont{Kransnyj}}, \bibnamefont{and}
  \bibinfo{author}{\bibfnamefont{P.}~\bibnamefont{Zoller}},
  \bibinfo{journal}{Eur. Phys. J. D} \textbf{\bibinfo{volume}{22}},
  \bibinfo{pages}{319} (\bibinfo{year}{2003}).

\bibitem[{\citenamefont{F{\"o}rstner et~al.}(2003)\citenamefont{F{\"o}rstner,
  Weber, Dankwerts, and Knorr}}]{foerstner:03a}
\bibinfo{author}{\bibfnamefont{J.}~\bibnamefont{F{\"o}rstner}},
  \bibinfo{author}{\bibfnamefont{C.}~\bibnamefont{Weber}},
  \bibinfo{author}{\bibfnamefont{J.}~\bibnamefont{Dankwerts}},
  \bibnamefont{and} \bibinfo{author}{\bibfnamefont{A.}~\bibnamefont{Knorr}},
  \bibinfo{journal}{Phys. Rev. Lett.} \textbf{\bibinfo{volume}{91}},
  \bibinfo{pages}{127401} (\bibinfo{year}{2003}).

\bibitem[{\citenamefont{Barenco et~al.}(1995)\citenamefont{Barenco, Deutsch,
  Ekert, and Josza}}]{barenco:95}
\bibinfo{author}{\bibfnamefont{A.}~\bibnamefont{Barenco}},
  \bibinfo{author}{\bibfnamefont{D.}~\bibnamefont{Deutsch}},
  \bibinfo{author}{\bibfnamefont{A.}~\bibnamefont{Ekert}}, \bibnamefont{and}
  \bibinfo{author}{\bibfnamefont{R.}~\bibnamefont{Josza}},
  \bibinfo{journal}{Phys. Rev. Lett.} \textbf{\bibinfo{volume}{74}},
  \bibinfo{pages}{4083} (\bibinfo{year}{1995}).

\bibitem[{\citenamefont{Troiani et~al.}(2000)\citenamefont{Troiani, Hohenester,
  and Molinari}}]{troiani.prb:00}
\bibinfo{author}{\bibfnamefont{F.}~\bibnamefont{Troiani}},
  \bibinfo{author}{\bibfnamefont{U.}~\bibnamefont{Hohenester}},
  \bibnamefont{and} \bibinfo{author}{\bibfnamefont{E.}~\bibnamefont{Molinari}},
  \bibinfo{journal}{Phys. Rev. B} \textbf{\bibinfo{volume}{62}},
  \bibinfo{pages}{R2263} (\bibinfo{year}{2000}).

\bibitem[{\citenamefont{Biolatti et~al.}(2000)\citenamefont{Biolatti, Iotti,
  Zanardi, and Rossi}}]{biolatti:00}
\bibinfo{author}{\bibfnamefont{E.}~\bibnamefont{Biolatti}},
  \bibinfo{author}{\bibfnamefont{R.~C.} \bibnamefont{Iotti}},
  \bibinfo{author}{\bibfnamefont{P.}~\bibnamefont{Zanardi}}, \bibnamefont{and}
  \bibinfo{author}{\bibfnamefont{F.}~\bibnamefont{Rossi}},
  \bibinfo{journal}{Phys. Rev. Lett.} \textbf{\bibinfo{volume}{85}},
  \bibinfo{pages}{5647} (\bibinfo{year}{2000}).

\bibitem[{\citenamefont{Zrenner et~al.}(2002)\citenamefont{Zrenner, Beham,
  Stufler, Findeis, Bichler, and Abstreiter}}]{zrenner:02}
\bibinfo{author}{\bibfnamefont{A.}~\bibnamefont{Zrenner}},
  \bibinfo{author}{\bibfnamefont{E.}~\bibnamefont{Beham}},
  \bibinfo{author}{\bibfnamefont{S.}~\bibnamefont{Stufler}},
  \bibinfo{author}{\bibfnamefont{F.}~\bibnamefont{Findeis}},
  \bibinfo{author}{\bibfnamefont{M.}~\bibnamefont{Bichler}}, \bibnamefont{and}
  \bibinfo{author}{\bibfnamefont{B.}~\bibnamefont{Abstreiter}},
  \bibinfo{journal}{Nature} \textbf{\bibinfo{volume}{418}},
  \bibinfo{pages}{612} (\bibinfo{year}{2002}).

\bibitem[{\citenamefont{Li et~al.}(2003)\citenamefont{Li, Wu, Steel, Gammon,
  Stievater, Katzer, Park, Piermarocchi, and Sham}}]{li:03}
\bibinfo{author}{\bibfnamefont{X.}~\bibnamefont{Li}},
  \bibinfo{author}{\bibfnamefont{Y.}~\bibnamefont{Wu}},
  \bibinfo{author}{\bibfnamefont{D.}~\bibnamefont{Steel}},
  \bibinfo{author}{\bibfnamefont{D.}~\bibnamefont{Gammon}},
  \bibinfo{author}{\bibfnamefont{T.~H.} \bibnamefont{Stievater}},
  \bibinfo{author}{\bibfnamefont{D.~S.} \bibnamefont{Katzer}},
  \bibinfo{author}{\bibfnamefont{D.}~\bibnamefont{Park}},
  \bibinfo{author}{\bibfnamefont{C.}~\bibnamefont{Piermarocchi}},
  \bibnamefont{and} \bibinfo{author}{\bibfnamefont{L.~J.} \bibnamefont{Sham}},
  \bibinfo{journal}{Science} \textbf{\bibinfo{volume}{301}},
  \bibinfo{pages}{809} (\bibinfo{year}{2003}).

\bibitem[{\citenamefont{Zurek}(2003)}]{zurek:03}
\bibinfo{author}{\bibfnamefont{W.~H.} \bibnamefont{Zurek}},
  \bibinfo{journal}{Rev. Mod. Phys.} \textbf{\bibinfo{volume}{75}},
  \bibinfo{pages}{715} (\bibinfo{year}{2003}).

\bibitem[{\citenamefont{Rabitz et~al.}(2000)\citenamefont{Rabitz, {de
  Vivie-Riedle}, Motzkus, and Kompka}}]{rabitz:00}
\bibinfo{author}{\bibfnamefont{H.}~\bibnamefont{Rabitz}},
  \bibinfo{author}{\bibfnamefont{R.}~\bibnamefont{{de Vivie-Riedle}}},
  \bibinfo{author}{\bibfnamefont{M.}~\bibnamefont{Motzkus}}, \bibnamefont{and}
  \bibinfo{author}{\bibfnamefont{K.}~\bibnamefont{Kompka}},
  \bibinfo{journal}{Science} \textbf{\bibinfo{volume}{288}},
  \bibinfo{pages}{824} (\bibinfo{year}{2000}).

\bibitem[{\citenamefont{Peirce et~al.}(1988)\citenamefont{Peirce, Dahleh, and
  Rabitz}}]{peirce:88}
\bibinfo{author}{\bibfnamefont{A.~P.} \bibnamefont{Peirce}},
  \bibinfo{author}{\bibfnamefont{M.~A.} \bibnamefont{Dahleh}},
  \bibnamefont{and} \bibinfo{author}{\bibfnamefont{H.}~\bibnamefont{Rabitz}},
  \bibinfo{journal}{Phys. Rev. A} \textbf{\bibinfo{volume}{37}},
  \bibinfo{pages}{4950} (\bibinfo{year}{1988}).

\bibitem[{\citenamefont{Borz{\`\i} et~al.}(2002)\citenamefont{Borz{\`\i},
  Stadler, and Hohenester}}]{borzi.pra:02}
\bibinfo{author}{\bibfnamefont{A.}~\bibnamefont{Borz{\`\i}}},
  \bibinfo{author}{\bibfnamefont{G.}~\bibnamefont{Stadler}}, \bibnamefont{and}
  \bibinfo{author}{\bibfnamefont{U.}~\bibnamefont{Hohenester}},
  \bibinfo{journal}{Phys. Rev. A} \textbf{\bibinfo{volume}{66}},
  \bibinfo{pages}{053811} (\bibinfo{year}{2002}).

\bibitem[{\citenamefont{Turukhin et~al.}(2002)\citenamefont{Turukhin,
  Sudarshanam, Shahriar, Musser, Ham, and Hemmer}}]{turukhin:02}
\bibinfo{author}{\bibfnamefont{A.~V.} \bibnamefont{Turukhin}},
  \bibinfo{author}{\bibfnamefont{V.~S.} \bibnamefont{Sudarshanam}},
  \bibinfo{author}{\bibfnamefont{M.~S.} \bibnamefont{Shahriar}},
  \bibinfo{author}{\bibfnamefont{J.~A.} \bibnamefont{Musser}},
  \bibinfo{author}{\bibfnamefont{B.~S.} \bibnamefont{Ham}}, \bibnamefont{and}
  \bibinfo{author}{\bibfnamefont{P.~R.} \bibnamefont{Hemmer}},
  \bibinfo{journal}{Phys. Rev. Lett.} \textbf{\bibinfo{volume}{88}},
  \bibinfo{pages}{023602} (\bibinfo{year}{2002}).

\bibitem[{\citenamefont{Bao et~al.}(2003)\citenamefont{Bao, Bragas, Furdyna,
  and Merlin}}]{bao:03}
\bibinfo{author}{\bibfnamefont{J.}~\bibnamefont{Bao}},
  \bibinfo{author}{\bibfnamefont{A.~V.} \bibnamefont{Bragas}},
  \bibinfo{author}{\bibfnamefont{J.~K.} \bibnamefont{Furdyna}},
  \bibnamefont{and} \bibinfo{author}{\bibfnamefont{R.}~\bibnamefont{Merlin}},
  \bibinfo{journal}{Nature Materials} \textbf{\bibinfo{volume}{2}},
  \bibinfo{pages}{175} (\bibinfo{year}{2003}).

\bibitem[{\citenamefont{Tamura}(1985)}]{tamura:85}
\bibinfo{author}{\bibfnamefont{S.}~\bibnamefont{Tamura}},
  \bibinfo{journal}{Phys. Rev. B} \textbf{\bibinfo{volume}{31}},
  \bibinfo{pages}{2574} (\bibinfo{year}{1985}).

\bibitem[{\citenamefont{Bartels et~al.}(1998)\citenamefont{Bartels, Dekorsky,
  Kurz, and K{\"o}hler}}]{bartels:98}
\bibinfo{author}{\bibfnamefont{A.}~\bibnamefont{Bartels}},
  \bibinfo{author}{\bibfnamefont{T.}~\bibnamefont{Dekorsky}},
  \bibinfo{author}{\bibfnamefont{H.}~\bibnamefont{Kurz}}, \bibnamefont{and}
  \bibinfo{author}{\bibfnamefont{K.}~\bibnamefont{K{\"o}hler}},
  \bibinfo{journal}{Appl. Phys. Lett.} \textbf{\bibinfo{volume}{72}},
  \bibinfo{pages}{2844} (\bibinfo{year}{1998}).

\bibitem[{\citenamefont{Kent et~al.}(2002)\citenamefont{Kent, Stanton, Challis,
  and Henini}}]{kent:02}
\bibinfo{author}{\bibfnamefont{A.~J.} \bibnamefont{Kent}},
  \bibinfo{author}{\bibfnamefont{N.~M.} \bibnamefont{Stanton}},
  \bibinfo{author}{\bibfnamefont{L.~J.} \bibnamefont{Challis}},
  \bibnamefont{and} \bibinfo{author}{\bibfnamefont{M.}~\bibnamefont{Henini}},
  \bibinfo{journal}{Appl. Phys. Lett.} \textbf{\bibinfo{volume}{81}},
  \bibinfo{pages}{3497} (\bibinfo{year}{2002}).

\bibitem[{\citenamefont{Mandel and Wolf}(1995)}]{mandel:95}
\bibinfo{author}{\bibfnamefont{L.}~\bibnamefont{Mandel}} \bibnamefont{and}
  \bibinfo{author}{\bibfnamefont{E.}~\bibnamefont{Wolf}},
  \emph{\bibinfo{title}{Optical coherence and quantum optics}}
  (\bibinfo{publisher}{Cambridge University Press},
  \bibinfo{address}{Cambridge}, \bibinfo{year}{1995}).

\bibitem[{\citenamefont{Viola and Lloyd}(1998)}]{viola:98}
\bibinfo{author}{\bibfnamefont{L.}~\bibnamefont{Viola}} \bibnamefont{and}
  \bibinfo{author}{\bibfnamefont{S.}~\bibnamefont{Lloyd}},
  \bibinfo{journal}{Phys. Rev. A} \textbf{\bibinfo{volume}{58}},
  \bibinfo{pages}{2733} (\bibinfo{year}{1998}).

\bibitem[{\citenamefont{McCall and Hahn}(1967)}]{mccall:67}
\bibinfo{author}{\bibfnamefont{S.~L.} \bibnamefont{McCall}} \bibnamefont{and}
  \bibinfo{author}{\bibfnamefont{E.~L.} \bibnamefont{Hahn}},
  \bibinfo{journal}{Phys. Rev. Lett.} \textbf{\bibinfo{volume}{18}},
  \bibinfo{pages}{908} (\bibinfo{year}{1967}).

\bibitem[{\citenamefont{Panzarini et~al.}(2002)\citenamefont{Panzarini,
  Hohenester, and Molinari}}]{panzarini.prb:02}
\bibinfo{author}{\bibfnamefont{G.}~\bibnamefont{Panzarini}},
  \bibinfo{author}{\bibfnamefont{U.}~\bibnamefont{Hohenester}},
  \bibnamefont{and} \bibinfo{author}{\bibfnamefont{E.}~\bibnamefont{Molinari}},
  \bibinfo{journal}{Phys. Rev. B} \textbf{\bibinfo{volume}{65}},
  \bibinfo{pages}{165322} (\bibinfo{year}{2002}).

\bibitem[{unp()}]{unpublished}
\bibinfo{note}{U. Hohenester and G. Stadler (unpublished).}

\bibitem[{\citenamefont{Bergmann et~al.}(1998)\citenamefont{Bergmann, Theuer,
  and Shore}}]{bergmann:98}
\bibinfo{author}{\bibfnamefont{K.}~\bibnamefont{Bergmann}},
  \bibinfo{author}{\bibfnamefont{H.}~\bibnamefont{Theuer}}, \bibnamefont{and}
  \bibinfo{author}{\bibfnamefont{B.~W.} \bibnamefont{Shore}},
  \bibinfo{journal}{Rev. Mod. Phys.} \textbf{\bibinfo{volume}{70}},
  \bibinfo{pages}{1003} (\bibinfo{year}{1998}).

\bibitem[{\citenamefont{Mandel et~al.}(2003)\citenamefont{Mandel, Greiner,
  Widera, Rom, H{\"a}nsch, and Bloch}}]{mandel:03}
\bibinfo{author}{\bibfnamefont{O.}~\bibnamefont{Mandel}},
  \bibinfo{author}{\bibfnamefont{M.}~\bibnamefont{Greiner}},
  \bibinfo{author}{\bibfnamefont{A.}~\bibnamefont{Widera}},
  \bibinfo{author}{\bibfnamefont{T.}~\bibnamefont{Rom}},
  \bibinfo{author}{\bibfnamefont{T.~W.} \bibnamefont{H{\"a}nsch}},
  \bibnamefont{and} \bibinfo{author}{\bibfnamefont{I.}~\bibnamefont{Bloch}},
  \bibinfo{journal}{Nature} \textbf{\bibinfo{volume}{425}},
  \bibinfo{pages}{937} (\bibinfo{year}{2003}).

\end{thebibliography}
\end{document}